\documentstyle[12pt]{article}
\makeatletter
\ifcase\@ptsize
 \font\teneufm=eufm10
 \font\seveneufm=eufm7
 \font\fiveeufm=eufm5
 \font\teneusm=eusm10
 \font\seveneusm=eusm7
 \font\fiveeusm=eusm5
\or
 \font\teneufm=eufm10 scaled \magstephalf
 \font\seveneufm=eufm7
 \font\fiveeufm=eufm5
 \font\teneusm=eusm10 scaled \magstephalf
 \font\seveneusm=eusm7
 \font\fiveeusm=eusm5
\or
 \font\teneufm=eufm10 scaled \magstep1
 \font\seveneufm=eufm7
 \font\fiveeufm=eufm5
 \font\teneusm=eusm10 scaled \magstep1
 \font\seveneusm=eusm7
 \font\fiveeusm=eusm5
\fi
\newfam\eufmfam
\newfam\eusmfam
\textfont\eufmfam=\teneufm  \scriptfont\eufmfam=\seveneufm
  \scriptscriptfont\eufmfam=\fiveeufm
\textfont\eusmfam=\teneusm  \scriptfont\eusmfam=\seveneusm
  \scriptscriptfont\eusmfam=\fiveeusm
\def\frak{\ifmmode\let\next\frak@\else
 \def\next{\errmessage{Use \string\frak\space only in math mode}}\fi\next}
\def\frak@#1{{\frak@@{#1}}}
\def\frak@@#1{\fam\eufmfam#1}

\def\sh{\ifmmode\let\next\sh@\else
 \def\next{\errmessage{Use \string\sh\space only in math mode}}\fi\next}
\def\sh@#1{{\sh@@{#1}}}
\def\sh@@#1{\fam\eusmfam#1}
\ifcase\@ptsize
 \font\tenmsa=msam10
 \font\sevenmsa=msam7
 \font\fivemsa=msam5
 \font\tenmsb=msbm10
 \font\sevenmsb=msbm7
 \font\fivemsb=msbm5
\or
 \font\tenmsa=msam10 scaled \magstephalf
 \font\sevenmsa=msam7
 \font\fivemsa=msam5
 \font\tenmsb=msbm10 scaled \magstephalf
 \font\sevenmsb=msbm7
 \font\fivemsb=msbm5
\or
 \font\tenmsa=msam10 scaled \magstep1
 \font\sevenmsa=msam7
 \font\fivemsa=msam5
 \font\tenmsb=msbm10 scaled \magstep1
 \font\sevenmsb=msbm7
 \font\fivemsb=msbm5
\fi
\newfam\msafam
\newfam\msbfam
\textfont\msafam=\tenmsa  \scriptfont\msafam=\sevenmsa
  \scriptscriptfont\msafam=\fivemsa
\textfont\msbfam=\tenmsb  \scriptfont\msbfam=\sevenmsb
  \scriptscriptfont\msbfam=\fivemsb
\def\Bbb{\ifmmode\let\next\Bbb@\else
 \def\next{\errmessage{Use \string\Bbb\space only in math mode}}\fi\next}
\def\Bbb@#1{{\Bbb@@{#1}}}
\def\Bbb@@#1{\fam\msbfam#1}
\def\hexnumber@#1{\ifnum#1<10 \number#1\else
 \ifnum#1=10 A\else\ifnum#1=11 B\else\ifnum#1=12 C\else
 \ifnum#1=13 D\else\ifnum#1=14 E\else\ifnum#1=15 F\fi\fi\fi\fi\fi\fi\fi}
\def\msa@{\hexnumber@\msafam}
\def\msb@{\hexnumber@\msbfam}
\mathchardef\square="0\msa@03
\makeatother
\newcommand{\beq}{\begin{equation}}
\newcommand{\eeq}{\end{equation}}
\newcommand{\ba}{\begin{array}}
\newcommand{\ea}{\end{array}}
\newcommand{\bea}{\begin{eqnarray}}
\newcommand{\eea}{\end{eqnarray}}
\newcommand{\bean}{\begin{eqnarray*}}
\newcommand{\eean}{\end{eqnarray*}}

\newtheorem{theorem}{Theorem}[section]
\newtheorem{prop}[theorem]{Proposition}
\newtheorem{lem}[theorem]{Lemma}

\newtheorem{remark}[theorem]{Remark}

\newtheorem{proof}{Proof.}
\newenvironment{pf}{\begin{proof} \rm}{\hfill$\square$ \end{proof}}
\newcommand{\CW}{{\cal W}}
\newcommand{\CP}{{\cal P}}

\newcommand{\CS}{{\cal S}}
\newcommand{\CN}{{\cal N}}
\newcommand{\CM}{{\cal M}}
\newcommand{\CQ}{{\cal Q}}
\newcommand{\CL}{{\cal L}}

\newcommand{\CV}{{\cal V}}
\newcommand{\CE}{{\cal E}}
\newcommand{\CX}{{\cal X}}
\newcommand{\CG}{{\cal G}}
\makeatletter
\@addtoreset{equation}{section}

\makeatother
\newcommand{\VV}{{\Bbb V}}
\newcommand{\WW}{{\Bbb W}}
         \def\Ga{\Gamma}
\def\be{\beta}

\def\la{\lambda}

\newcommand{\cmp}[3]{Comm. Math. Phys. {\bf #1} (#2), #3}

\newcommand{\plb}[3]{Phys. Lett. {\bf B #1} (#2), #3}

\newcommand{\faa}[3]{Funct. Anal. Appl. {\bf #1} (#2), #3}

\newcommand{\lmp}[3]{Lett. Math. Phys. {\bf #1} (#2), #3}

\newcommand{\rmp}[3]{Rev. Math. Phys. {\bf #1} (#2), #3}

\newcommand{\rref}[1]{(\ref{#1})} %puts parentheses around ref's

\newcommand{\third}{\frac{1}{3}}
\newcommand{\twth}{\frac{2}{3}}

\def\dsl{\displaystyle}

\newcommand{\del}{{\partial}}

\def\rigau#1#2{\mathop{\buildrel{\dsl#1}\over{\hbox
to#2pt{\rightarrowfill}}}}

\def\dpt#1#2{\frac{\partial #1}{\partial t_{#2}}}

\def\la{\lambda}                                  
                                  
\def\Alg{{\frak G}}                                
                              
\def\alg{{\frak g}}                               
                             
\def\fraksl{{\frak s}{\frak l}}

\def\im{\mbox{Im}}             
\def\Ker{\mbox{Ker}}

\def\res{\mbox{res}}

\def\var{manifold}                        
                
\def\tenp{Poisson tensor}                 
\def\pare{bracket}                     
                       
\def\varp{Poisson \var}                
\def\varb{\bih\ \var}                  
         
\def\camp{vector field}                
  
\def\bih{bihamiltonian}                
\def\ham{Hamiltonian}

\def\parn{\par\noindent}

\def\nota#1 {{\bigskip\centerline{*********}
\par \noindent {#1}\par \centerline{*********}\bigskip}}

\def\dpt#1#2{\frac{\partial #1}{\partial t_{#2}}}

\def\linendpf{\hfill$\square$\par\medskip\par\noindent}

\def\Alg{{\frak G}}

\def\alg{{\frak g}}

\def\var{manifold}
\begin{document}
\baselineskip=18pt
\begin{titlepage}
\begin{flushright}
Ref. SISSA 72/97/FM
\end{flushright}
\vspace{0.8truecm}
\begin{center}
{\huge Bihamiltonian Reductions \\ \vspace{.5truecm}
and ${\CW}_n$--Algebras}
\end{center}
\vspace{0.8truecm}
\begin{center}
{\large
Paolo Casati${}^1$, Gregorio Falqui${}^2$,\\
Franco Magri${}^1$, and
Marco Pedroni${}^3$}\\
\vspace{1.truecm}
${}^1$ Dipartimento di Matematica, Universit\`a di Milano\\
Via C. Saldini 50, I-20133 Milano, Italy\\
E--mail: casati@vmimat.mat.unimi.it,
magri@vmimat.mat.unimi.it\\
${}^2$ SISSA, Via Beirut 2/4, I-34014 Trieste, Italy\\
E--mail: falqui@sissa.it\\
${}^3$ Dipartimento di Matematica, Universit\`a di Genova\\
Via Dodecaneso 35, I-16146 Genova, Italy\\
E--mail: pedroni@dima.unige.it
\end{center}
\vspace{0.2truecm}
\begin{abstract}{\noindent 
We discuss the geometry of the Marsden--Ratiu
reduction theorem for a bihamiltonian manifold.
We consider the case of  the manifolds 
associated with the Gel'fand--Dickey 
theory, i.e., loop algebras over $\fraksl_n$.
We provide an explicit identification,
tailored on the MR reduction, of
the Adler--Gel'fand--Dickey brackets
with the Poisson brackets on the 
reduced bihamiltonian manifold $\CN$.
Such an identification relies 
on a suitable immersion of $T^*\CN$ into the algebra of 
pseudo differential operators connected to geometrical
features of the theory of (classical) $\CW_n$--algebras}.
\end{abstract}
%\vspace{1.truecm}
\begin{center}
%{\today}
\end{center}
\vspace{1.truecm}
Work supported by the Italian M.U.R.S.T. and by
the G.N.F.M. of the Italian C.N.R. 
\end{titlepage}
\setcounter{footnote}{0}
\section{Introduction}
\label{sec1}
$\CW$--algebras are 
algebras with {\em quadratic} commutation relations
admitting the Virasoro Lie Algebra
%\[
%[l_n,l_m]=(m-n)l_{n+m} + c (n^3-n) \delta_{m+n,0}
%\]
as a subalgebra. They
have been the object of extensive study in the last few years, 
after the identification of such a structure (due
to  A. B. Zamolodchikov~\cite{Za85})
as the extended symmetry algebras of relevant models
of two--dimensional (Quantum) Conformal Field Theory.
It was soon understood  that
a physically meaningful family of  
such algebras could be obtained as quantum deformations 
of the  Adler--Gel'fand--Dickey (AGD)
bracket well known from the classical theory of soliton 
equations~\cite{FaLu90,Lu88,FeFr91,Ba89,DIZ91}.
Such a discovery prompted a remarkable amount of 
work aiming both at the 
classification of {\em all possible} $\CW$--algebras, and the  
study of their representations and geometrical aspects 
(see, for an updated review,~\cite{Wrev}).
A particularly fruitful approach proved to be 
the Drinfel'd--Sokolov (DS) reduction scheme for 
Kac--Moody Lie algebras~\cite{DS}, for the wide variety of examples
it embodies and the appealing mathematics behind it (see, 
e.g.,~\cite{Kho87,FeFr90,Fehetal92,Fraetal93,BGHM93}.) \par
In this paper we want to discuss some of the geometrical aspects
of the theory of (classical) $\CW$--algebras related to the Hamiltonian
%and bihamiltonian theory of 
approach to infinite--dimensional integrable systems.
We set our study in the framework of the 
Marsden Ratiu (MR) reduction
scheme~\cite{MR} for Poisson manifolds, extended in~\cite{CMP}
to the case of \bih\ \var s, whose application to the
Gel'fand--Dickey hierarchies \cite{GD} can be found in~\cite{CMP,CFMP3}.\par
The MR and the DS reduction schemes
can be compared as follows, in relation to the 
theory of ``Hamiltonian systems 
with symmetry'' (see, e.g.,~\cite{AM}). The basic datum of
the DS scheme is a Poisson action of a group $G$ 
on Poisson manifold $\CM$. 
The group defines a 
momentum map and a Hamiltonian reduction of the manifold $\CM$ (the 
{\em Marsden--Weinstein} reduction). The reduction process 
considers a submanifold ${\CS_G}$ of 
$\CM$ (a level surface of the momentum 
map), a foliation ${\CE_G}$ of ${\CS_G}$ 
(the orbits of the little group),
and the reduced phase space $\CN={\CS_G}/{\CE_G}$.
On the other hand, in the geometric scheme of the bihamiltonian
MR reduction the two steps are defined in terms of {\em two}   
compatible Poisson brackets on $\CM$.
The submanifold $\CS$ is a symplectic leaf of the first Poisson bracket,
and a foliation $\CE$ is generated on it 
by the restriction 
to $T\CS$ of the image, via the second bracket, of  
the Casimir functions of the first one.
The resulting quotient space $\CN=\CS/\CE$ is a bihamiltonian manifold.
\par
The first point of this paper (Section \ref{sec2})
is to discuss in some detail the 
interplay of the manifolds $\CM$, $\CS$, and $\CN$
together with their Poisson structures. We pay special attention
to the Poisson brackets {\em on 1--forms}. This fits
the results of~\cite{Kho87} that identify classical
$\CW_n$ algebras as Poisson algebras of 1--forms over 
GD manifolds of monic differential operators of order $n+1$.
\par
From Section~\ref{sec3} onwards, 
we specialize our constructions to the case of loop
algebras $L(\fraksl_n)$
over the simple Lie algebra $\alg=\fraksl_n$, which we 
equip with the compatible Poisson structures:
\begin{equation}
\begin{array}{l}
P_0:=\mbox{the commutator with a fixed element } A\in L(\fraksl_n);\\
P_1:=\mbox{the modified Kirillov--Kostant one.}
\end{array}\nonumber
\end{equation} 
In Section~\ref{sec4} and~\ref{sec5}
we  identify the reduced Poisson brackets
with the well known linear and quadratic 
AGD brackets on the algebra of pseudo differential operators (see,
e.g.,~\cite{A,D}), and provide the concrete examples
of the KdV and Boussinesq cases.\par 
It is worthwhile to remark that there are several  ways to provide 
the phase space of the Gel'fand--Dickey theories a (bi)hamiltonian 
structure, and namely:
\par
1) the Adler--Gel'fand--Dickey procedure~\cite{D,Ma79}; \par
%which will be briefly resumed in Section~\ref{secsix}
2) the DS reduction from the space of matrix
valued first order differential operators;\par
3) the bihamiltonian reduction process on a loop algebra.\\
The equivalence between {1)} and {2)} is one of the results
of Drinfel'd--Sokolov seminal paper~\cite{DS}, while the one between
{3)} and {2)} has been proved %(by one of the present authors) 
in~\cite{Pe95}, using the fact that the Marsden--Weinstein 
symmetry reduction can be seen as a particular case
of the Marsden--Ratiu reduction.
This paper provides a direct and {\em constructive}
proof of the equivalence between the AGD
brackets and those obtained by Hamiltonian reduction 
in the MR framework. Its main 
aim is the description of the geometrical properties 
of the MR reduction processes.
Accordingly, we explicitly construct 
the embedding of $T^*\CN$ into the algebra of 
pseudodifferential operators which allows us to perform 
the identification between the MR--reduced Poisson brackets
and the AGD ones.
In particular, we make contact with 
the results of~\cite{Ra91} (see also~\cite{Dico93})
about the factorization
of the AGD quadratic Poisson tensor into a pair of 
Poisson algebra homomorphisms. 
We show indeed that the latter are a particular 
instance of the Poisson morphisms  discussed in Section~\ref{sec2}. 
Since each of those mappings corresponds to a 
precise step in the MR reduction process,
our results may provide the theory of the $\CW_n$--algebras
some further geometrical flavour.
\section{The Marsden--Ratiu  Reduction}\label{sec2}
In this section we will briefly describe the bihamiltonian reduction
process developed in \cite{CMP}, which is based on the
Marsden--Ratiu reduction theorem \cite{MR}.
Let us first recall the few notions  of the general
theory of Poisson and bihamiltonian manifolds which are
needed for our purposes.\par
A {\it Poisson manifold\/} is a manifold $\CM$ endowed with a Poisson
bracket $\{\cdot,\cdot\}$, i.e., a bilinear skewsymmetric composition
law of $C^\infty$--functions fulfilling the Leibniz rule
and the Jacobi identity. The corresponding {\it
Poisson tensor} $P$ is the  bivector field $P$ 
on $\CM$, considered as a linear skewsymmetric map $P:T^*\CM\to
T\CM$, defined by 
\beq
\{f,g\}=\langle df,Pdg\rangle.\label{P}
\eeq
Any Poisson bracket on 
functions induces a Poisson 
bracket on forms \cite{Do,MaMo84}. 
If $\alpha_1$ and $\alpha_2$ are arbitrary
1--forms on a Poisson manifold $\CM$ and $P$ is the Poisson tensor, 
the bracket $\{\alpha_1,\alpha_2\}_P$ is defined by
its value on a 
vector field $X$ by 
\beq
\left<\{\alpha_1,\alpha_2\}_P,X\right>=L_{P\alpha_1}
\left<\alpha_2,X\right> -L_{P\alpha_2}
\left<\alpha_1,X\right>-\left<\alpha_1,L_X(P)\alpha_2\right>,\label{pf}
\eeq
where $L_{P\alpha_1}\left<\alpha_2,X\right>$
denotes the Lie derivative 
of the scalar function $\left<\alpha_2,X\right>$ along the
vector field $P\alpha_1$, and $L_X(P)$ is the Lie derivative
 of the Poisson tensor $P$ along $X$.\par
A {\it bihamiltonian manifold\/} is a manifold endowed 
with a pair of compatible Poisson brackets $\{\cdot, \cdot\}_0$ 
and $\{\cdot,\cdot\}_1$. Two Poisson brackets are compatible if the 
linear combinations
\beq
\{f,g\}_\lambda:=\{f,g\}_1+\lambda \{f,g\}_0\label{pp}
\eeq
verify the Jacobi identity for any value of the parameter $\lambda$.
This is tantamount to requiring that the cyclic compatibility condition
\beq
\begin{array}{rl} \{f,\{g,h\}_0\}_1&+\{h,\{f,g\}_0\}_1+\{g,\{h,f\}_0\}_1+\\
+&\{f,\{g,h\}_1\}_0+\{h,\{f,g\}_1\}_0+\{g,\{h,f\}_1\}_0=
0\end{array}\label{cp}
\eeq
holds for any triple of functions $(f,g,h)$  on $\CM$.
In this case, the
bracket $\{\cdot,\cdot\}_\lambda$ is called the {\it Poisson pencil} on $\CM$
 defined by $\{\cdot,\cdot\}_0$ and $\{\cdot,\cdot\}_1$.\par
In order to describe the bihamiltonian reduction process, let us
recall that every \varp\ $(\CM,P)$ is foliated in symplectic leaves.
Indeed, the characteristic distribution $C=\{Pdf\mid f\in C^\infty(\CM)\}$ is
integrable, and the maximal integral leaves are symplectic           
\var s \cite{LM}. %The Casimir functions of $P$ are constant on $C$. 
On a bihamiltonian manifold the symplectic leaves of both
Poisson tensors can be further foliated. Let us
denote by $C$ the characteristic distribution of the
Poisson tensor $P_0$. A second distribution $D$, defined by
$D=\{P_1df\vert\ f \mbox{ is a Casimir function of\ }  P_0\}$ 
is naturally conjugated to $C$.  
The reduction theory of \varb s is the study of the interplay 
between these two distributions.
As a consequence of the compatibility condition~\rref{cp} 
between the Poisson brackets, 
the distribution $D$ is integrable \cite{CMP}.         
Let us choose a specific symplectic leaf $\CS$ of $\{\cdot,\cdot\}_0$,
and let us denote by $E$ the distribution induced on $\CS$ by $D$; 
thus  the
leaves of $E$ are the intersections of $\CS$ with the leaves of $D$.
We shall assume $E$ sufficiently regular that there exists the
quotient space $\CN=\CS/E$, and  denote by $i_{\CS}:\CS\to \CM$ 
and $\pi:\CS\to \CN$
the canonical immersion of $\CS$ in $\CM$ and the canonical projection
of $\CS$ onto $\CN$. Then~\cite{CMP}:
\begin{prop} The quotient space $\CN=\CS/E$ is a bihamiltonian
manifold. On $\CN$ there exists a unique Poisson pencil
$\{\cdot,\cdot\}^\CN_\lambda$ such that
\beq
\{f,g\}^\CN_\lambda\circ\pi=\{F,G\}_\lambda\circ i_{\CS}\label{pn}
\eeq
for any pair of functions $F$ and $G$ which extend the functions $f$
and $g$ of $\CN$ into $\CM$, and are constant on $D$. Technically, this
means that the function $F$ verifies the conditions
\bea 
F\circ i_{\CS}=f\circ\pi\label{fn}\\
\{F,K\}_1=0\label{vv}
\eea
for any function $K$ whose differential, at the points of 
$\CS$, belongs to the kernel of $P_0$.\end{prop}
%\endpf
Since in this paper we will deal with \tenp s
rather than  \pare s, it is worthwhile to discuss the meaning of
the previous proposition in terms of \tenp s. 
First of all we prove the following
\begin{lem}\label{tangS}
Let $s\in \CS$ and $v\in T^*_s\CM$. Then $v$ is in the 
annihilator  $(D^0)_s$ of $D$ at $s$
if and only if $(P_\la)_s v$ is tangent to $\CS$.
\end{lem}
{\bf Proof.} $(P_\la)_s v$ is tangent to $\CS$  
if and only if $\langle w, (P_\la)_s v\rangle=0$ for all
$w \in \Ker(P_0)_s$. But this is equivalent to the statement that
$\langle v, (P_1)_s w\rangle=0$ for all
$w \in \Ker(P_0)_s$, i.e., that $v \in (D^0)_s$.
\linendpf
To construct the reduced Poisson pencil $P^\CN_\la$ starting from
the Poisson pencil $P_\la$ on $\CM$ we have to
conform to the following scheme:\par
\begin{itemize}
\item[1.] For any 1--form $\alpha$ on $\CN$, and we consider
the 1--form $\pi^*\alpha$ on $\CS$, which obviously belongs to
the annihilator $E^0$ of $E$ in $T^*\CS$   since $E=\Ker \pi_*$.  
\item[2.] We construct a {\em lifting} of $\alpha$, that is, 
a 1--form $\beta$ on $\CM$ which belongs to the annihilator $D^0$ of $D$ 
and satisfies 
\begin{equation}
i_\CS^*\beta = \pi^*\alpha.\label{D01}
\end{equation}
Such a lifting $\beta$ of $\alpha$ is not uniquely defined,
but this arbitrariness is irrelevant. 

%Let us remark that the conditions ~\rref{D01}--\rref{D0} 
%do not define uniquely the 1--form $\beta$. There are, indeed, 
%infinitely many  $\beta$ corresponding to the same 1--form $\alpha\in 
%T^*\CN$, and they form an equivalence class. However, they have
%the same restriction $\pi^*\alpha \in E^0$ to $\CS$.
\item[3.] We construct the vector field $P_\la\beta$ 
associated with the  1--form $\beta$ through the Poisson 
pencil of $\CM$. Thanks to Lemma~\ref{tangS}, we have that 
$P_\la\beta$ is tangent to $\CS$.
\item[4.] Finally, we project this vector field from $\CS$ to $\CN$. 
The projection of $P_\la \beta$ does not
depend on the choice of the particular lifting $\beta$ 
and  defines unambiguously $(P_\la^\CN)\alpha$.
\end{itemize}
We notice that
in the construction of the reduced pencil
$P^\CN_\la$  only the value of the 1--form $\beta$ at
the points of $\CS$ plays a role.
Moreover, even if $\CS$ is {\em not} a \bih\ \var, to any lifting
$\beta$ of a 1--form on $\CN$, we can associate a whole pencil of
vector fields $X_\la=P_\la(\be)$ {\em tangent} to $\CS$.\par
%Let $i_\CS^*(T^*\CM)$ be the pull-back of the cotangent
%bundle $T^*\CM$ on $\CS$. 
Let us denote by $\CX(\CM)$ (resp. $\CX^*(\CM)$) the space of vector fields 
(resp. 1--forms) on $\CM$, by $\Gamma(\CV)$ the space of sections of 
the bundle $\CV\to \CM$,  and by  $i_\CS^*(\CV)$ its pull-back on $\CS$. 
In the space of sections of $i_\CS^*(T^*\CM)$ we can
define the subspace $\Gamma_{\CS}$ formed by all sections of $i_\CS^*(D^0)$
whose restriction to $\CX(\CS)$ are liftings of 1--forms  on $\CN$: 
\beq
\Gamma_{\CS}=\{\beta \in \Gamma(i^*_{\CS}(D^0))|\ \beta_{\vert 
\CX(\CS)}\in \im \pi^*\}.
\eeq
By definition of $\Gamma_\CS$, a surjective map 
$J_\CS:\Gamma_\CS\to \CX ^*(\CN)$ can be introduced by means of
\beq
\pi^* ({J_\CS} (\beta)) =i_\CS^*(\beta).\label{mor0}
\eeq
Then the previous scheme about the definition of the reduced
Poisson pencil can be summarized in the formula 
\beq
P^\CN_\la \circ {J_\CS}=\pi_*\circ P_\lambda.\label{t1}
\eeq
\begin{prop} \label{mor1} The space $\Gamma_{\CS}$ is closed with 
respect to the Poisson brackets on 1--forms
$\{\cdot,\cdot,\}_{P_\la}$. Moreover the map $J_\CS$ 
is a morphism between the Lie algebras
$(\Gamma_\CS,\{\cdot,\cdot\}_{P_\la})$ and 
$(\CX^*(\CN),\{\cdot,\cdot\}_{P_\la^{\CN}})$.
\end{prop}
{\bf Proof.} First of all we remark that, 
since for all $\beta \in \Gamma_\CS$ the vector field
$P_\la \beta$ is tangent to $\CS$, the right--hand side of the 
expression
\beq\label{1003}
\left<\{\beta_1,\beta_2\}_{P_\la},X\right>=L_{P_\la\beta_1}
\left<\beta_2,X\right> 
-L_{P_\la\beta_2}\left<\beta_1,X\right>-
\left<\beta_1,L_X(P_\la)\beta_2\right>,
\eeq
where $\beta_1$, $\beta_2$ $\in \Gamma_{\CS}$, defines 
a section of $i_\CS^*(T^*\CM)$. In order to prove that
$\{\beta_1,\beta_2\}_{P_\la}\in \Gamma_{\CS}$ we have first 
to check that, for all $Y$ in $D$,
\beq
\langle \{\beta_1,\beta_2\}_{P_\la},Y\rangle=0.\label{D00}
\eeq
Since $D$ is generated by the Hamiltonian vector
fields $P_1df$ with 
Hamiltonians $f$ which are Casimirs of $P_0$, we can consider 
$Y=P_1df$ with $P_0 df=0$. Being $\beta_1,\beta_2 \in \Gamma_{\CS}$,
we have
$L_{P_\la\beta_2}\langle \beta_1,Y\rangle=
L_{P_\la\beta_1}\langle \beta_2,Y\rangle=0$;
hence we simply have to prove that 
$\langle \beta_1,L_Y(P_\la)\beta_2\rangle=0$.
Actually this is a consequence of the following formula (equivalent to
the compatibility condition \rref{cp}):
\beq
L_{P_1df}(P_0)+L_{P_0df}(P_1)=0\qquad \forall f \in C^\infty(\CM).
\eeq
We are thus left with showing that the Poisson bracket
$\{\beta_1,\beta_2\}_{P_\la}$ restricted to $\CX(\CS)$ is in the 
image of $\pi^*$.
We will show both this property and that $J_\CS$ is a morphism 
by proving that 
\beq
\langle \{\beta_1,\beta_2\}_{P_\la},X\rangle=\langle
\{J_\CS(\beta_1),J_\CS(\beta_2)\}_{P_\la^\CN},\pi_*(X)\rangle\label{rf}
\eeq
holds true 
for any vector field $X$ tangent to $\CS$ and 
projectable onto $\CN$. 
From~\rref{1003} and  the identity 
$\langle \beta_1,L_X(P_\la)\beta_2)\rangle
=\langle J_\CS(\beta_1),L_{\pi_*(X)}(P_\la^\CN) J_\CS(\beta_2)\rangle$,
which can be proved using properties of the Lie
derivative and relation \rref{t1}, we have:
\beq
\begin{array}{l}
\langle \{\beta_1,\beta_2\}_{P_\la},X\rangle=L_{P_\la\beta_1}
\langle \beta_2,X\rangle-L_{P_\la\beta_2}\langle \beta_1,X\rangle 
-\langle \beta_1,L_X(P_\la)\beta_2\rangle\\ \qquad
=\frac{\del}{\del t_1} \langle J_\CS(\beta_2),\pi_* (X)\rangle
-\frac{\del}{\del t_2}\langle J_\CS(\beta_1),\pi_* (X)\rangle 
-\langle J_\CS(\beta_1),L_{\pi_* (X)}(P^\CN_\la)J_\CS(\beta_2)\rangle\\ \qquad
=\langle \{J_\CS(\beta_1),J_\CS(\beta_2)\}_{P_\la^\CN},\pi_*
(X)\rangle,
\end{array}
\eeq
where we have denoted  the Lie derivative along
${P^\CN_\la}(J_\CS(\beta_i))$ by $\frac{\del}{\del 
t_i}$. 
\hfill$\square$\par\noindent
\subsection{The transversal submanifold}\label{sec31}
In this subsection we recall the  technique of the
{\em transversal submanifold\/} described in~\cite{CP}. 
Besides allowing to
simplify the calculation involved
in the bihamiltonian reduction scheme, it  will naturally 
give rise to another space of sections $\Gamma_\CQ$  %\subset \CX^*(\CM)$ 
which is closed with respect the Poisson bracket on
1--forms, and to a Poisson morphism $J_\CQ\colon \Ga_\CQ\to \CX^*
(\CN)$ which, under an additional assumption, becomes an
isomorphism.\par
In the notations of the previous section,
a transversal submanifold to the distribution $E$ is a 
submanifold $\CQ$ of $\CS$, which intersects every integral leaves of
the distribution $E$ in one and only one point. 
This condition implies the following relation on the tangent spaces:
\beq
 T_Q\CS=T_Q\CQ\oplus E_q\qquad\forall\ Q\in \CQ.\label{tra}
\eeq
If  such a transversal submanifold $\CQ$ exists, then it is obviously 
diffeomorphic to the quotient manifold $\CN$, and inherits from $\CN$ 
a bihamiltonian structure. The Poisson pencil on $\CN$ can be computed 
by noticing~\cite{CP} 
that  given a 1--form $\alpha\in \CX^*(\CN)$ 
there always exists  a section   $\beta$ of $i^*_{\CQ}(T^*\CM)$ 
where $i_\CQ:\CQ\to \CM$ is the canonical immersion, such that 
\bea
P_\la \beta&\in& \CX(\CQ) \label{lq}\\
\langle \beta,Y\rangle&=&\langle \alpha,\pi^\CQ_*Y\rangle 
\qquad \forall Y\in \CX(\CQ),\label{tra1}
\eea
where $\pi^\CQ$ is the projection from $\CQ$ to $\CN$, i.e., the restriction
of $\pi$ to $\CQ$.
Therefore in order to compute the action of the reduced Poisson pencil
$P_\la^\CN$ on the 1--form  $\alpha$, 
one has simply to determine the expression of $P_\la \beta$.
\par
The sections   $\beta$ satisfying~\rref{lq}  
form a subset $\Gamma_\CQ$ of sections of ${i}_\CQ^*(T^*
\CM)$, and
a natural map $J_\CQ:\Gamma_\CQ\to \CX^*(\CN)$ is defined. Moreover, if 
$\beta\in \Gamma_{\CQ}$, then the value $\beta_Q$ 
of $\beta$ at $Q$ belongs to $(D^0)_Q$ for all
$Q\in \CQ$ (again by Lemma~\ref{tangS}).
%, since $P_\la \beta\in T\CQ$ implies $\beta\in D^0$.
%Indeed if $P_\la \beta$ belongs to $\CX(\CQ)$ then trivially 
%$(P_\la\beta)_Q$ 
%is in $T_Q\CS,\;\forall \la.$ Therefore 
%one has
%\beq
%\langle \beta,P_1\eta \rangle=-\langle P_1 \beta,\eta\rangle=0\qquad
%\forall \eta \in \Ker P_0.
%\eeq
From~\rref{pf} it can be seen that the bracket 
$\{\beta_1,\beta_2\}_{P_\la}$ of two elements of $\Gamma_\CQ$ is
a well--defined element of $\Gamma_\CQ$. Indeed,
to check this, it is
enough to show that $P_\la \{\beta_1,\beta_2\}_{P_\la}\in 
\CX(\CQ)$. Since $\CQ$ is a
submanifold, this follows from the relation (see \cite{Do}): 
\beq
P_\la\{\beta_1,\beta_2\}_{P_\la} =[P_\la \beta_1,P_\la
\beta_2].\label{pbf} 
\eeq
 In the same way  it can be proved that 
the   map $J_\CQ$ owns the same properties of the map $J_\CS$, i.e., 
$J_\CQ$ is a Lie algebra morphism from $\Gamma_\CQ$ to $\CX^*(\CN)$ 
and satisfies the relation 
\beq
\pi^\CQ_* \circ P_\la=P^\CN_\la\circ J_\CQ.\label{t112}
\eeq
Finally we observe that if
the kernels of the Poisson tensors $P_0$ and $P_1$ 
have trivial intersection on $\CQ$, then the map $J_{\CQ}$ 
becomes a Poisson isomorphism. It holds, indeed, the following 
\begin{prop} If $\Ker (P_0)\cap\Ker (P_1)=\{0\}$ on $\CQ$, then
for all $\alpha \in \CX^*(\CN)$ there exists a unique lifting
$\beta \in \Gamma_{\CQ}$ satisfying 
conditions~\rref{lq}--\rref{tra1}. 
Therefore $J_\CQ$  is an isomorphism.
\end{prop}
{\bf Proof.} We have only to prove that $J_{\CQ}$ is injective. 
Let us therefore suppose that there exist $\beta_1$ and $\beta_2$ 
in $\Gamma_\CQ$  such
that $J_{\CQ}(\beta_1)=J_{\CQ}(\beta_2)$. Then, using equation~\rref{t112}, 
$\pi_*(P_\la(\beta_1-\beta_2))=P_\la^\CN(J_{\CQ}(\beta_1-\beta_2))=0$.
Since  $P_\la(\beta_1-\beta_2)\in \CX(\CQ)$ and $\CQ$ is a transversal
submanifold we have that $P_\la(\beta_1-\beta_2)=0$ $\forall \la$.
%$\forall\la$ , we have $\pi_*(P_\la(\beta_1-\beta_2))=0$ $\forall \la$. Thus
%$P_\la(\beta_1-\beta_2)=0$ $\forall \la$, because
%$P_\la(\beta_1-\beta_2)\in \CX(\CQ)$ and $\CQ$ is a transversal
%submanifold. But this last relation 
This implies $\beta_1-\beta_2\in \Ker
(P_0)\cap  \Ker (P_1)=\{0\}$.
\hfill$\square$\par\noindent
\section{The Gel'fand--Dickey manifolds}\label{sec3}
In this  section we will specialize the Marsden--Ratiu reduction scheme 
to the class of bihamiltonian manifolds which correspond to
the Gel'fand--Dickey theories and their associated 
$\CW_n$--algebras~\cite{DS,FaLu90,FeFr90,Kho87}. 
%The geometrical setting for such theories is a 
%special class of bihamiltonian manifolds constructed over Lie 
%algebras. 
Let ${\frak g}$ be the simple Lie  algebra ${\frak s}{\frak l} (n+1)$ 
and $\CM={\frak G}$ the space of $C^\infty$--maps from $S^1$ into $\frak g$.
We denote by $x$ the coordinate on $S^1$, and by $S$ a map from 
$S^1$ into $\frak g$.
An element in the tangent space $T_S{\CM}$ is denoted by $\dot S$. 
%The cotangent space is i with the 
Identifying $\alg$ with $\alg^*$ by means of the Killing  form
$(\cdot,\cdot)$, a covector is  a map $V$ 
from $S^1$ into $\frak g$, whose value on the tangent vector $\dot S$  is
given by
\beq
\langle V,\dot S\rangle=\int_{S^1}(V(x),{\dot S}(x))dx
.\label{bl}
\eeq
The first Poisson bracket on $\CM$ is defined by
\beq
\{ f,g\}_0=-\langle A,[df(S),dg(S)]\rangle,\label{bp0}
\eeq
where $f$ and $g$ are arbitrary 
functionals on $\CM$, $df(S)$ and
$dg(S)$ are their differentials and 
$A$ is the vector of minimal weight for the ``usual'' Cartan 
decomposition of $\frak g$, i.e.,
\beq
A=\left(\begin{array}{ccccc} 
        0 & 0 &\dots & 0\\
           \dots &\dots & \dots & \dots\\
            \dots &\dots & \dots & \dots\\
           0 & 0 &\dots & 0\\
           1 & 0 &\dots & 0\end{array}\right).\label{a}
\eeq
The second Poisson bracket is defined by
\beq\{ f,g\}_1={\sigma}(df,dg)+\langle S,[df(S),dg(S)]\rangle,\label{bp1}
\eeq
where 
\beq
{\sigma}({\dot S}_1,{\dot S}_2)=\int_{S^1}({\dot S}_1,\frac{d}{dx}
{\dot S}_2)
dx\label{co}
\eeq
is the nontrivial cocycle on $\CM$.  
The corresponding Poisson tensors are given by
\bea
(P_0)_SV&=&[A,V]\label{p00}\\
(P_1)_SV&=&V_x +[V,S]\label{p01}
\eea
where $V_x$ denotes the derivative of the loop $V$ with respect to $x$.
The {\it Lie--Poisson pencil\/} $P_\la$ is 
\beq
(P_\la)_SV=V_x +[V,S+\la A].\label{2.16}
\eeq
It is a standard result \cite{LM} that these brackets are compatible.\\
%for any choice of $A$.
The reduction process starts with the choice of a 
specific symplectic leaf $\CS$
of the Poisson tensor \rref{p00}.
%\beq
%{\dot S}=[A,V].\label{4.5}
%\eeq 
All leaves are affine hyperplanes modeled on the orthogonal space 
${\frak G}^\perp_A$ (with respect to the pairing \rref{bl}) of 
${\frak G}_A=\{V\in {\frak G}\mbox{ s.t. } [V,A]=0\}$, and we choose 
the one passing through the sum $B$ of the root vectors  
corresponding to the positive simple roots of ${\frak g}$. 
%point 
%\beq
%B=\left(
%\begin{array}{ccccc} 
%        0 & 1 & \dots & \dots & 0\\
%         0 & 0 & 1 & \dots & 0\\
%         \dots & \dots & \dots & \dots & \dots \\
%         0 & \dots & \dots & \dots & 1\\
%         0 & \dots & \dots & 0 & 0 \end{array}\right).\label{4.6}
%\eeq
Thus 
$S\in \CS$ is parameterized by $2n$ periodic functions $(p_a,q_a)$
as follows
\beq
S=
\left(\begin{array}{ccccc} 
        {p}_0 & 1 & \dots & \dots & 0\\
         {p}_1 & 0 & 1 & \dots & 0\\
         \dots & \dots & \dots & \dots & \dots \\
         p_{n-1} & \dots & \dots & \dots & 1\\
         {q}_0 & \dots & \dots & {q}_{n-1} & -{p}_0\end{array}
\right).\label{sym}
\eeq
The Gel'fand--Dickey  theories are related to the particular choice of
the pair $(A,B)$ above made but it is possible 
to make different choices of such elements. In \cite{CFMP5}
were considered pairs corresponding to the ``fractional'' (or generalized)
KdV hierarchies~\cite{Be91,BGHM93} and in~\cite{MoPi95} those leading
to the classical AKNS system. \par
The next step is the study  of  the foliation $E$.
A basic feature of (generalized) GD theories  
is that, thanks to the form of the Kirillov-Kostant
Poisson tensor~\rref{2.16}, the integral leaves of $E$ are orbits of a 
group action. 
From~\cite{CP,CFMP2} we borrow 
\begin{prop} The following properties hold:
\begin{description}
\item[{i)}] The distribution $E$ is spanned by the vector fields
\beq
X_V(S)=V_x+[V,S]\label{4.9}
\eeq
with $V\in\Alg_{AB}=\left\{ V\in {\frak G}_A\ |\ V_x+[V,B]\in {\frak
G}_A^\perp\right\}.$
\item[{ii)}] The integral leaves of $E$ are the orbits of the gauge action of 
${\CG}_{AB}=\exp({\frak G}_{AB})$ on $\CS$ defined by 
\beq
S'=G S G^{-1}+G_x G^{-1}.\label{4.10}
\eeq
\item[{iii)}] The points of $\CS$ with coordinates
$p_j=0$ form a submanifold $\CQ$ of $\CS$ transversal to the
distribution $E$.
The reduced bihamiltonian Gel'fand--Dickey manifold $\CN = \CS/E$
is therefore parametrized by $n$ independent functions  
$(u_0,u_1,\dots, u_{n-1})$ on $\CS$, invariant under ${\CG}_{AB}$.
The restriction to $\CQ$ of the projection $\pi:\CS\to \CN=\CS/E$. 
is simply given by the
equations 
\beq
u_j=q_j\qquad (j=0,1,\dots,n-1).\label{tq0}
\eeq
\end{description}
\end{prop}
To compute the reduced Poisson pencil $P_\la^{\CN}$
according to the general scheme discussed in Section~\ref{sec2}, 
we will  use the technique of the transversal manifold 
outlined in  subsection~\ref{sec31}.
It is worthwhile to remark that
the points $Q$ of $\CQ$ have the canonical Frobenius form 
\beq
Q=\left(
\begin{array}{ccccc} 
        0 & 1 & \dots & \dots & 0\\
         0 & 0 & 1 & \dots & 0\\
         \dots & \dots & \dots & \dots & \dots \\
         0 & \dots & \dots & \dots & 1\\
         u_0 & u_1 & \dots & u_{n-1} & 0 \end{array}\right).\label{tq}
\eeq
\begin{prop}
On the whole symplectic leaf $\CS$ one has that   
$\Ker (P_0)\cap \Ker (P_1)=\{0\}.$
\end{prop}
{\bf Proof.} The statement is proved recalling that the Lie algebra $\alg$
admits the gradation
\beq
\fraksl_{n+1}=\oplus_{k=-n}^{n} \alg_k,
\eeq
where $\alg_k$ is the space of matrices $M$ such that $M_j^l=0$ for
$l-j\neq k$. This induces a corresponding
grading in the loop algebra $\Alg$. Let us now consider an
element $V\in \Ker P_0$. Then its decomposition with respect to
the gradation is 
\beq
V=\sum_{k=-n}^{n-1} V_k
\eeq
and each homogeneous component $V_k$ is in ${\frak G}_A$.
Remark that any element of the symplectic leaf $\CS$ 
can be decomposed as $S=T+B$ with $T\in \oplus_{k=-n}^{0} \alg_k$. 
Imposing that $V\in \Ker P_1$ and considering the maximal degree
elements, we get $[B,V_{n-1}]=0$. Hence, $V_{n-1}$ commutes both 
with $A$ and $B$. Then, since $\alg_A\cap\alg_B=\{0\}$ (see~\cite{CP}),
we can conclude that $V_{n-1}=0$. In the same way, a recursive argument
proves that $V=0$. \hfill$\square$\par\medskip\par\noindent
In particular, the statement of this proposition holds
at the points of  $\CQ$, so that, recalling the discussion
of Section~\ref{sec31}, the morphism $J_{\CQ}$ is an isomorphism, whose 
inverse will be denoted by $\phi$. \par
%Let us therefore compute  
The unique section  $\VV=\phi(v)$ in $\Gamma_{\CQ}$ 
lifting  the 1--form $v$ 
in $\CX^*(\CN)$ can be explicitly computed, using conditions~\rref{lq}
and ~\rref{tra1}.
Let $v=(v_0,v_1,\dots,v_{n-1}) \in \CX^*(\CN)$
be 
defined by $\langle v, \dot u \rangle =\sum_{i=0}^{n-1}\int_{S^1} v_i\dot u_i$
and let $(\VV^j_i)_{i,j=0,\dots,n}$ be the entries of the matrix
$\VV=\phi(v)$; then \rref{tra1} implies
\beq
\VV^j_n=v_j\qquad j=0,\dots,n-1.\label{li}
\eeq
Substituting the explicit expression of the Poisson pencil
\rref{2.16} in \rref{lq} we get
\beq
\VV_x+ [\VV,Q+\la A]\in T_Q\CQ.\label{pqt}
\eeq
This condition implies the following relations on the entries of the
matrix $\VV$:
\beq
\begin{array}{cccccc}
-\VV^{k+1}_0&&+\VV^k_{0x}&+&\VV^k_n(u_0+\la)&=0\\
-\VV^{k+1}_1+&\VV^k_0+&  \VV^k_{1x}&+&\VV^k_nu_1&=0\\
-\VV^{k+1}_2+&\VV^k_1+&\VV^k_{2x}&+&\VV^k_nu_2&=0\\
\dots&&\dots &&\dots&\\
-\VV^{k+1}_n&+\VV^k_{n-1}&+&\VV^k_{nx}&&=0. 
\end{array}\label{5.9}
\eeq
These formulas together with the zero trace condition 
$\sum_{k=0}^n\VV_k^k=0$ show that each of the elements  $\VV^k_l$ is obtained 
algebraically from
the first $n$ elements of the last column of $\VV$. 
%In our particular choice of
%the lifting of the 1--form $v=(v_0,\dots,v_{n-1})$ the significative 
%invariant of the lifting process are the entries of 
%the first row of $\VV$, since the
%entries of the last column just coincides with the coordinate of $v$.\par
With the elements of $\Gamma_{\CQ}$  at our disposal, the reduction 
of the Poisson pencil can be now completed quite easily. According to 
the general procedure we have to compute the vector field
\beq
\dot Q=\VV_x+[\VV,Q+\la A],\label{5.21}
\eeq
where $\VV=\phi(v)\in \Gamma_\CQ$.
By expanding equation \rref{5.21} we obtain
\beq
\begin{array}{l}
{\dot
u}_0=(\VV^n_0)_x+\VV^n_n(u_0+\la)
-\sum_{l=0}^{n-1}u_l{\VV}_0^l-\la{\VV}_0^0\\
{\dot
u}_j=(\VV^n_j)_x+{\VV}_{j-1}^{n}+{\VV}_n^n u_j
-\sum_{l=0}^{n-1}u_l{\VV}_j^l
-\la{\VV}_j^0\qquad j=1,\dots n-1\label{5.24}
\end{array}
\eeq 
and 
\beq
0=(\VV^n_n)_x+{\VV}_{n-1}^{n}-\sum_{l=0}^{n-1}u_l{\VV}_n^l
-\la{\VV}_n^0.\label{upun}
\eeq 
Equations \rref{5.24} give
the explicit form of the reduced Poisson pencil.
In order to obtain a more compact formula, to be used in the following 
section, we define
\beq
\label{extev}
\begin{array}{l}
\VV^{n+1}_0=\VV^n_{0x}+\VV^n_n(u_0+\la)\\
\VV^{n+1}_l=\VV^n_{lx}+\VV^n_{l-1}+u_l\VV^n_n\qquad  l=1,\dots,n,
\end{array}
\eeq
where $u_n:=0$.
Then the reduced Poisson pencil \rref{5.24} can be written as
\beq
\label{redpoibis}
{\dot u}_j=\VV_j^{n+1}-\sum_{l=0}^{n-1}u_l\VV_j^l-\la\VV^0_j
\qquad  l=0,\dots,n-1.
\eeq
\section{The Adler--Gel'fand--Dickey brackets}\label{sec4}
In this section we perform the identification of the reduced 
brackets on $\CN$ with the AGD brackets 
and discuss 
%make contact with the 
%results of~\cite{Ra91} concerning 
some \ham\ aspects of Gel'fand--Dickey 
theories and ${\cal W}_n$--algebras.\par
The usual setting for Gel'fand--Dickey theories can be briefly
described as follows. One considers 
the space $\Psi DO$  of  pseudodifferential operators on $S^1$
(see, e.g.,~\cite{A,DJKM,D,Ma79}
for a broader account on the subject),
i.e., the space of formal Laurent series of the form
\beq
A=\sum_{i=-\infty}^N a_i(x)\del_x^i.
\eeq
It is an associative algebra under the 
product defined (on homogeneous
elements $A_1= a_{k_1} \del_x^{k_1},\; A_2= a_{k_2} \del_x^{k_2}$) as
\beq
A_1 A_2=\sum_{n\ge 0} {{n}\choose{k_1}} a_1 (\del_x^k a_2) \cdot 
\del_x^{k_2+k_1-n}.
\eeq
Its associated Lie algebra admits a filtration (indexed by the 
integers) via the
subspaces $\Psi DO_{p}$ formed by those operators of order at most $p$,
and a trace form (the {\em Adler trace}) given by
\beq
tr(A)= \oint \mbox{res}_\del\; A=\oint a_{-1}(x).
\eeq 
It is customary to denote
by $A_+$ the strictly differential part of $A$ and by $A_-:=A-A_+$.\par
Let $\CL_{n+1}$ be the space of order $n+1$ monic differential operators
on the circle, parametrized by the $(n+1)$--tuple of functions 
$\{u_0,\ldots,u_{n}\}$. Its tangent space $T\CL_{n+1}$ can be naturally 
identified with the space $D_{n}$ of differential operators
of order $n$, and (via the Adler trace) its cotangent space 
$T^*\CL_{n+1}$ with the quotient  space $\Psi DO/\Psi DO_{-(n+2)}$ 
of pseudodifferential operators modulo those of degree less than $-(n+1)$.
It is a classical result that $\CL_{n+1}$ is endowed with
a compatible pair of Poisson tensor, which are defined by
\begin{equation}\label{4.4}
\begin{array}{l}
\CP_0(L)\cdot X= [X,L]_+\\
\CP_1(L)\cdot X= L (XL)_+-(LX)_+L,
\end{array}
\eeq  
and are usually called, respectively,
the Gardner--Zakharov--Faddeev and the Adler--Gel'fand--Dickey brackets
or collectively AGD brackets. It is also a well-known fact (see, 
e.g.,~\cite{D}) that these brackets restrict to the subspace $u_n=0$,
where $L={\partial}^{n+1}-\sum_{j=0}^{n-1}u_j{\partial}^j$. \\
The connection of our previous results  with this picture 
is established simply by translating into this 
language  the two processes 
involved in the MR reduction:
\begin{description}
\item [{1.}] the lifting of 1--forms from $\CX^*(\CN)$ in
$\Gamma_{\CQ}$;
\item [{2.}] the projection of the \ham\ \camp s from $\CS$ to $\CN$.
\end{description}
This can be done as follows.
With any $u\in \CN$, ${\dot u}\in \CX(\CN)$, $v\in \CX^*(\CN)$, and $\VV
\in \Gamma_\CQ$
we associate the  operators $(L,{\dot L},
\xi(v),\psi(\VV))$ 
defined by
\beq
\begin{array}{rl}\label{6.1}
L&:={\partial}^{n+1}-\sum_{j=0}^{n-1}u_j{\partial}^j\\
\dot L&:=-\sum_{j=0}^{n-1}{\dot u}_j\del^j\\
\xi(v)&%:=-\sum_{j=0}^{n}{\partial}^{-j-1}\VV^j_n
:=-\sum_{j=0}^n{\del}^{-j-1}[\phi(v)]_n^j\\
\psi(\VV)&:=-\sum_{j=0}^n\VV_j^0\del^j.
\end{array}
\eeq
Notice that 
$\VV_j^0$ are the elements of the  first row of $\VV$ and
$[\phi(v)]_n^j$ are the elements of the last column 
of the image under $\phi$ of $v$.
For simplicity of notation, and to anticipate the content of the next
subsection, we put $X=\xi(v)$ and $E=\psi(\VV)$.
Since $[\phi(v)]^j_n=v_j$, for $j=0,\dots,n-1$, we have
\beq
\int_{S^1}\res_\del(X\dot L)\,dx=\langle v,\dot u\rangle,
\label{6.3}
\eeq
and so we are allowed to consider  $(L,X)$ as 
the representative of the 
1--form $v$ in the space of pseudodifferential operators. 
Our task is to establish a link between $E$ and $X$, corresponding
to the relation between $v$ and the matrix $\VV$ constructed in the
previous section, and a link between  
$E$ and $\dot L$, corresponding
to the relation between the 1--form $\VV$ and the vector field ${\dot 
u}$ given by formula~\rref{redpoibis}. The key to solve this problem is
given by 
\begin{lem} \label{4.1} Let the coefficients $\VV^j_l$ be defined, for 
$j=0,\dots,n+1$ and $l=0,\dots,n$, by equations 
\rref{5.9} and \rref{extev}. Then
\beq
\del^jE=-\sum_{l=0}^n\VV^j_l\del^l+(\del^jEL^{-1})_+(L-\la),\qquad 
j=0,\dots,n+1.
\label{6.6}
\eeq
\end{lem}
{\bf Proof.} We prove by induction that 
\beq
\del^jE=-\sum_{l=0}^n\VV^j_l\del^l+R_j(L-\la),
%\qquad j=0,\dots,n+1,
\label{6.7}
\eeq
where $R_j$ is a purely differential operator.
It is true for 
$j=0$ because of the definition of $E$. Moreover
\beq
\begin{array}{rl}
\del^{j+1}E&=\del(\del^jE)=-\sum_{l=0}^n({\VV^j_l}_x\del^l+\VV^j_l\del^{l
+1})+\del R_j(L-\la)\\
&=-\sum_{l=0}^n({\VV^j_l}_x+\VV^j_{l-1})\del^l-\VV^j_n\del^{n+1}+
\del R_j(L-\la)\\
&=-\sum_{l=0}^n({\VV^j_l}_x+\VV^j_{l-1})\del^l-\VV^j_n(L+\sum_{l=0}^{n-
1}u_l\del^l)+
\del R_j(L-\la)\\
&=-\sum_{l=0}^n({\VV^j_l}_x+\VV^j_{l-1}+u_l\VV^j_n)\del^l-\VV^j_nL
+\del R_j(L-\la),
\end{array}
\eeq
where we put $u_n=0$ and $\VV^j_l=0$ for $l<0$. Now we use the 
recursion relations \rref{5.9} and the definition of $\VV^{n+1}_j$
to get 
\beq
\begin{array}{rl}
\del^{j+1}E&=
-\sum_{l=0}^n\VV^{j+1}_l\del^l+\la \VV^j_n-\VV^j_nL
+\del R_j(L-\la)\\
&=-\sum_{l=0}^n\VV^{j+1}_l\del^l+(\del R_j-\VV^j_n)
(L-\la),
\end{array}
\eeq
proving \rref{6.7}. Finally we notice that this equation implies 
that 
\beq
R_j=\left[(\del^jE)(L-\la)^{-1}\right]_+=
\left[(\del^jE)(L^{-1}+\la L^{-2}+\dots)\right]_+=
(\del^jEL^{-1})_+
\eeq
for $j=0,\dots,n+1$.
\hfill$\square$\par\noindent
We are now able to prove the main relations 
between the pseudodifferential 
operators associated with our geometrical objects. From now on we put 
$\VV=\phi(v)$ and, consequently, $E=\psi(\phi(v))$.
\begin{prop}\label{5.3}\null\par\noindent
\begin{description}
\item [{i)}] The operators $E$ and $X$ 
are related by $E=(XL)_+$.
This equation is the operator form of the ``lifting of covectors''
entering the MR reduction. 
\item [{ii)}] The operators $E$ and $\frac{dL}{dt_i}$ associated with 
the i--th reduced structure are related by
\beq
\begin{array}{rl}
\displaystyle{\frac{dL}{dt_1}}&=LE-(LEL^{-1})_+L\label{6.8}\\
\displaystyle{\frac{dL}{dt_0}}&=E-(LEL^{-1})_+.\label{6.9}
\end{array}
\eeq
These equations are the operator form of the projection of the 
Hamiltonian vector fields from $\CS$ onto $\CN$.
\item[{iii)}] The operator form of the reduced Poisson tensor on
$\CN$ is given by
\beq 
\begin{array}{rl}
\displaystyle{\frac{dL}{dt_1}}&=L(XL)_+ - (LX)_+L\\
\displaystyle{\frac{dL}{dt_0}}&=(XL)_+ - (LX)_+=[X,L]_+,
\end{array}\label{6.10}
\eeq
i.e.,
the reduced brackets 
are the Adler--Gel'fand--Dickey brackets in the algebra 
of pseudodifferential operators.
\end{description}
\end{prop}
{\bf Proof.}\parn
i) It is easily seen that the first claim is equivalent to the 
assertion that $EL^{-1}=X+Z$, with $\deg Z=-n-2$. The latter follows 
from Lemma~\ref{4.1}, since 
\beq
\res(\del^jEL^{-1})=\res\left(-\sum_{l=0}^n\VV^j_l\del^lL^{-1}+(\del^jX
)_+-\la (\del^jX)_+L^{-1}\right),
\eeq
and $\deg(\del^jX)_+L^{-1}\leq -2$ for $j\leq n$. 
Therefore $\res(\del^jEL^{-1})=-\res\sum_{l=0}^n\VV^j_l\del^lL^{-1}=
-\res(\VV^j_n\del^nL^{-1})=-\VV^j_n$ 
for $j=0,\dots,n$. We conclude that $EL^{-1}=-(\del^{-1}\VV_n^0+\dots+
\del^{-(n+1)}\VV^n_n)+Z$, with $\deg Z=-n-2$. Finally, $\sum_{j=1}^n
(-\del^{-j-1}\VV^j_n)=X$ from \rref{li}.\parn
ii) Let us set $\dot L=\frac{dL}{dt_1}-\la\frac{dL}{dt_0}$;
then $\dot L=-\sum_{l=0}^{n-1}{\dot u}_j\del^j$, where ${\dot u}_j$ is 
the vector field associated with the Poisson pencil \rref{redpoibis}:
\beq
{\dot u}_j=\VV_j^{n+1}-\sum_{l=0}^{n-1}u_l\VV_j^l-\la\VV^0_j.
\eeq
Then 
\beq
\dot L=\sum_{j=0}^{n-1}\left(-\VV_j^{n+1}+
\sum_{l=0}^{n-1}u_l\VV_j^l+\la\VV^0_j\right)\del^j
=\sum_{j=0}^n\left(-\VV_j^{n+1}+\sum_{l=0}^{n-1}u_l\VV_j^l+
\la\VV^0_j\right)\del^j,
\eeq
since equations \rref{upun} and the definition \rref{extev} of 
$\VV_n^{n+1}$ imply that
$\VV_n^{n+1}-\sum_{l=0}^{n-1}u_l\VV_n^l-\la\VV^0_n=0$. 
Therefore we have 
\beq
\begin{array}{rl}
\dot L&=-\sum_{j=0}^n\VV_j^{n+1}\del^j+\sum_{l=0}^{n-1}u_l\sum_{j=
0}^n\VV^l_j\del^j+\la \sum_{j=0}^n\VV^0_j\del^j\\
&=\del^{n+1}E-(\del^{n+1}EL^{-1})_+(L-\la)+\sum_{l=0}^{n-1}u_l\left(
-\del^{l}E+(\del^{l}EL^{-1})_+(L-\la)\right)-\la E\\
&=(\del^{n+1}-\sum_{l=0}^{n-1}u_l\del^{l})E-\left[
(\del^{n+1}EL^{-1})_+-\sum_{l=0}^{n-1}u_l(\del^{l}EL^{-1})_+\right]L\\
&\phantom{=}-\la \left(E+\sum_{l=0}^{n-1}u_l(\del^{l}EL^{-1})_+
-(\del^{n+1}EL^{-1})_+\right)\\
&=LE+RL-\la (E-(LEL^{-1})_+),
\end{array}
\eeq
where $R$ is a purely differential operator.
Thus 
\beq
\begin{array}{rl}
\frac{dL}{dt_1}&=LE+RL\\
\frac{dL}{dt_0}&=-E+(LEL^{-1})_+,
\end{array}
\eeq
and
\beq
\frac{dL}{dt_1}L^{-1}=LEL^{-1}+R\Longrightarrow R=-(LEL^{-1})_+.
\eeq
By  this result, one easily gets
\beq
\begin{array}{rl}
\displaystyle{\frac{dL}{dt_1}}&=L(XL)_+ -(LX)_+L\label{6.10b}\\
\displaystyle{\frac{dL}{dt_0}}&=(XL)_+-(LX)_+=[X,L]_+,\label{6.11b}
\end{array}
\eeq
so that iii) is also proved. \hfill$\square$\par\noindent
%The link between the pseudo--differential approach and the
%Marsden--Ratiu reduction procedure
%leads to 
\subsection{On Radul's morphism}
We are now in a position to analyse from a geometrical point of view
some results of~\cite{Ra91} (furtherly generalized in~\cite{FiRa93}).
They can be described as follows.\\
Let $\CE$ be the algebra of differential operators 
with coefficients in the space of differential
polynomials in the $u_i$'s and ${\CE}_{n+1}$ its quotient
with respect to the equivalence relation
\beq
\label{radequi}
E\sim F \mbox{ iff } F = E+ RL \mbox{ for some } R\in \CE.
\eeq
Moreover, let $W_L$ be the map associating with every element $E$ of $\CE$
the differential operator 
\beq
W_L(E):=LE-(LEL^{-1})_+L\equiv (LEL^{-1})_- L.
\eeq  
Finally, define $\Theta_L(E):= (EL^{-1})_-$ and $\Phi_L(X):=(XL)_+.$
\begin{prop}
The following properties hold~\cite{Ra91}:
\begin{itemize}
\item [i)]  $W_L$ takes values in  $T\CL_{n+1}$, 
passes to the quotient ${\CE}_{n+1}={\CE}/\sim$, 
and is an isomorphism of Lie algebras, provided one defines on ${\CE}_{n+1}$
the commutator
\beq
[E,F]_L:=[E,F] + W_L(E)\cdot (F)-W_L(F)\cdot (E)\mbox{ mod }
\sim.\label{Rad}
\eeq 
Here $W_L(Y)\cdot (F)=\frac{d}{d\epsilon}\vert_{\epsilon = 0}
F(L+\epsilon W_L(Y))$;
\item [ii)]
the map $W_L$ factorizes as
\beq
W_L= \CP_1\cdot \Theta_L,
\eeq
where $\CP_1$ is the Poisson tensor defined in equation~\rref{4.4};
\item [iii)]
$\Theta_L$ is a  Lie algebras isomorphism between 
$({\CE}_{n+1},[\cdot,\cdot]_L)$ and 
$(\CX^*(\CL_{n+1}),\{\cdot,\cdot\}_{\CP_1})$, where
$\{\cdot,\cdot\}_{\CP_1}$ is the Poisson bracket on 1--forms 
induced by $\CP_1$.
\end{itemize}
\end{prop}
Let us regard $\CE_{n+1}$ as a fiber bundle over $\CL_{n+1}$, and the
map $\psi$ of equation~\rref{6.1} as a map from $\Gamma_\CQ$ to 
$\Gamma(\CE_{n+1})$.
We endow the space $\Gamma_{\CQ}$  with the Poisson bracket on
1--forms corresponding to the Poisson tensor $P_1$ of equation~\rref{p01}, 
which can be shown to have the form
\beq
\{V_1,V_2\}_1=
\frac{\partial V_2}{\partial t_1}-\frac{\partial V_1}{\partial t_2}
-[V_1,V_2],\label{5.16}
\eeq
where ${\del}\over {\del {t}_i}$ is the directional 
derivative along the vector 
field $P_1 V_i$ and $[\cdot,\cdot]$ is the matrix commutator.
%The map $\psi$ has the following property: 
%becomes a Lie algebra morphism, as stated in
\begin{prop}\label{5.4} The map $\psi$ 
is a Lie algebra homomorphism between 
$(\Gamma_{\CQ},\{\cdot,\cdot\}_1)$ and  $(\CE_{n+1} , [\cdot,\cdot]_L)$.
\end{prop}
{\bf Proof.}
Let $\VV$, $\WW$ $\in \Gamma_\CQ$, and let us put
$E=\psi(\VV)=-\sum_{j=0}^n\VV^0_j\del^j$ and $F=\psi(\WW)
=-\sum_{j=0}^n\WW^0_j\del^j$; then from formula~\rref{5.16} we have
\beq
\psi(\{\VV,\WW\}_1)=\frac {\partial F}{\partial t_1}
-\frac {\partial E}{\partial t_2}-\psi([\VV,\WW]),
\eeq
where ${{\del}\over {\del {t}_1}}$ and ${{\del}\over {\del
{t}_2}}$ are the directional derivatives along the vector 
fields $P_1 \VV$ and  $P_1 \WW$, respectively.
It follows from Proposition \ref{5.3} that 
%the Poisson structures
%on $\CX^*(\CL_{n+1})$ and $\CX^*(\CN)$ coincide, so that 
$P_1\VV$ (resp.\ $P_1\WW$) is, in terms of pseudodifferential 
operators, $W_L(E)$ (resp.\ $W_L(F)$)). Therefore we have 
\beq
\frac{\del F}{\del t_1}-\frac{\del E}{\del t_2}=W_L(E)(F)-W_L(F)(E).
\eeq
Therefore it remains to prove that
\beq\label{4.29}
-\psi([\VV,\WW])=\sum_{l,j=0}^n (\VV^0_j\WW^j_l-\WW^0_j\VV^j_l)\del^l
\sim [E,F]. 
\eeq
Using formula~\rref{6.7} with $\la=0$, we have
\beq
\begin{array}{rl}
[E,F]&=EF-FE\\
&=-\sum_{j=0}^n\VV^0_j\del^jF +\sum_{j=0}^n\WW^0_j\del^j E\\
&=\sum_{l,j=0}^n (\VV^0_j\WW^j_l-\WW^0_j\VV^j_l)\del^l+ RL,
\end{array}
\eeq
where $R$ is a differential operator. Hence~\rref{4.29} follows,
and the proof is completed.
\linendpf
The results of this section can be summarized 
in the following commutative diagram of Lie algebra morphism 
\begin{equation}\label{diagramma}
\begin{array}{ccc}
({ {\CE}}_{n+1},[\cdot,\cdot]_L)&\rigau{{\Theta}_L}{50}&(\CX^*({\CL}_{n+1}),
\{\cdot,\cdot\}_{\CP_1})\\
{\psi}{\Bigg\uparrow}&&{\Bigg\uparrow}{\xi}\\
(\Gamma_{\CQ},\{\cdot,\cdot\}_{P_1})&\rigau{J_{\CQ}}{50}&
(\CX^*(\CN),\{\cdot,\cdot\}_{P_1^{\CN}})
\end{array}\nonumber
\end{equation}
Notice that  $\xi$  is indeed a morphism since it allowed us 
to identify the AGD brackets and the reduced brackets 
in Proposition~\ref{5.3}. Hence, the fact that $\Theta_L$ is a 
morphism of Lie algebras can be read as a consequence of the
general properties of $J_\CQ$ discussed in subsection~\ref{sec31}.

\section{Examples}\label{sec5}
This last section is devoted to exemplify the
results of this paper in the cases of the KdV and Boussinesq
hierarchies. \par
The KdV hierarchy is the GD theory associated with the loop algebra
$L({\fraksl} (2))$ whose picture in the Marsden--Ratiu scheme
has been given in~\cite{CMP}. 
As we have seen in Section~\ref{sec31}, to compute
the reduced Poisson pencil and the maps $\xi$ and $\psi$ it is
enough to calculate the elements of the subspace $\Gamma_{\CQ}$,
where  the transversal manifold $\CQ$
is parametrized by
\beq
Q=\left(
\begin{array}{cc} 
        0 & 1 \\
         u & 0 \end{array}\right).\label{tq1}
\eeq
The lifting $\VV$ in $\Gamma_{\CQ}$ of the 1-form $v$ of
$\CX^*(\CN)$ is determined by the conditions
\rref{li} and \rref{5.9}, 
and has the form
\beq
\VV=\left(
\begin{array}{cc} 
-\frac{1}{2}v_x\         & v \\
      -\frac{1}{2}v_{xx}+(u+\la)v    & \frac{1}{2}v_x
 \end{array}\right).\label{vv3}
\eeq
The reduced Poisson pencil is given by formula~\rref{5.24}:
\beq
{\dot u}=-\frac{1}{2} v_{xxx}+2(u+\la)v_x +u_{x} v.\label{rp}
\eeq
The Poisson brackets on 1--forms is
\beq
\{v_1,v_2\}_{P_\la^\CN}=\dpt{v_2}{1}-\dpt{v_1}{2}-[v_1,v_2]_\CN,
\eeq
where $\dpt{}{i}$ is the directional derivative along the vector 
field $P_\la^\CN v_i$, and 
\beq
[v_1,v_2]_\CN=v_1v_{2x}-v_2v_{1x}.
\eeq
The Lax operator $L$ is defined by 
\beq
L={\del }^2-u,
\eeq
while the pseudodifferential operators $X$ and $E$ can be read off
the  matrix $\VV$ and are
\bea
X=\xi(v)&=&-{\del }^{-1}v-{\del }^{-2}
(\frac{1}{2}v_x)\label{x1}\\ 
E=\psi(\phi(v))&=&\frac{1}{2}v_x-v{\del }.\label{psi1}
\eea
The relations $E=(XL)_+$ and the
commutativity of the diagram~\rref{diagramma} 
are easily checked.\par\bigskip
The next example  is the Boussinesq hierarchy,
which is associated with the loop algebra
over ${\fraksl}(3)$. Again we simply have to compute 
the elements of $\Gamma_{\CQ}$, where the transversal submanifold
$\CQ$ is given by
\beq
Q=\left(
\begin{array}{ccc} 
        0 & 1 & 0\\
         0 & 0 & 1\\
         u_0 & u_1 & 0 \end{array}\right).\label{tq2}
\eeq
Given the 1--form $v=(v_0,v_1)$ in $\CX^*(\CN)$, the matrix $\VV=\phi(v)$
is uniquely characterized by
conditions \rref{li} and \rref{5.9}, together with the obvious zero trace
relation, to be 
\beq \VV=
\left(
\begin{array}{ccc}
\twth v_{0xx}-\twth v_0 u_1-v_{1x} & v_1-v_{0x}&v_0\\
&&\\
\begin{array}{c}\twth v_{0xxx}-\twth v_{0x}u_1-v_{1xx}+\\
(u_0-\twth u_{1x}+\la ) v_0\end{array}
&-\third v_{0xx}+\third v_0 u_1 & v_1\\
&&\\ 
\VV^{2}_0&  \VV^{2}_1 & -\third v_{0xx}+\third v_0 u_1+v_{1x}
\end{array}
\right),\label{vvv1}
\eeq
where
\beq\begin{array}{rl}
\VV^{2}_0&=- {v}_{1xxx}+({u}_{0} - {\frac {4}{3}}{u}_{1x}+\lambda)
{v}_{0x} +({u}_{0x}    - 
{\frac {2}{3}}{u}_{1xx}) {v}_{0} - {\frac {2}{3}}{v}_{0xx} {{u}_{1}} +
\\& ({{u}_{0}}+\la) {v}_{1}+{ \frac {2}{3}} {v}_{0xxxx}
\\
\VV^{2}_1&=-v_{1xx}+\third v_{0xxx}-\third v_{0x}u_1+
(u_0-\third u_{1x}+\la) v_0 +v_1u_1.
\end{array}
\eeq
The reduced Poisson pencil is
\beq
\begin{array}{rl}
{\dot u}_0=& 
  \frac23v_{0xxxxx} 
- \frac43u_1v_{0xxx} 
- 2u_{1x}v_{0xx} 
+ (\frac23u_1^2 + 2u_{0x} - 2u_{1xx})v_{0x}\\
& + (- \frac23u_{1xxx}+ \frac23u_1u_{1x}+ u_{0xx})v_0
- v_{1xxxx} 
+ u_1v_{1xx} 
+ 3u_0v_{1x} 
+ u_{1x}v_1 
+ 3\la\, v_{1x},\\
{\dot u}_1=&
  v_{0xxxx}
- u_1v_{0xx}
+ (3u_0 - 2u_{1x})v_{0x} 
+ (2u_{0x}- u_{1xx})v_0\\
&- 2v_{1xxx} 
+ 2u_1v_{1x}
+ u_{1x}v_1 
+ 3\la\, v_{0x}.
\end{array}\label{sr2}
\eeq
In this case the 
operators $L$, $X$ and $E$ are
\beq
\begin{array}{l}
L={\del }^3-u_1 {\del }-u_0\\
X=-{\del }^{-1}v_0-{\del }^{-2}v_1-{\del }^{-3}(v_{1x}-\third v_{0xx}
+\third v_0 u_1)\\ 
E=-(\twth v_{0xx}-\twth v_0 u_1-v_{1x})-(v_1-v_{0x}){\del }-v_0{\del }^{2}.
\end{array}
\eeq

\end{document}